# Copper and Transparent-Conductor Reflectarray Elements on Thin-Film Solar Cell Panels

Philippe Dreyer, Monica Morales-Masis, Sylvain Nicolay, Christophe Ballif and Julien Perruisseau-Carrier, *Senior Member, IEEE*

*Abstract*—This work addresses the integration of reflectarray antennas (RA) on thin film Solar Cell (SC) panels, as a mean to save real estate, weight, or cost in platforms such as satellites or transportable autonomous antenna systems. Our goal is to design a good RA unit cell in terms of phase response and bandwidth, while simultaneously achieving high optical transparency and low microwave loss, to preserve good SC and RA energy efficiencies, respectively. Since there is a trade-off between the optical transparency and microwave surface conductivity of a conductor, here both standard copper and transparent conductors are considered. The results obtained at the unit cell level demonstrates the feasibility of integrating RA on a thin-film SC, preserving for the first time good performance in terms of both SC and RA efficiency. For instance, measurement at X-band demonstrate families of cells providing a phase range larger than 270° with average microwave loss of -2.45dB (resp. -0.25dB) and average optical transparency in the visible spectrum of 90% (resp. 85%) using transparent conductive multilayer (resp. a copper layer).

## I. INTRODUCTION

RECENT years have witnessed growing interest in the integration of antennas into Solar Cell (SC) to gain space, weight, visual disturbance, or costs [1-8]. In space applications for instance, both solar panels and communication system are major contributors to the overall size and weight of the satellites and combining these two systems could save real estate and cost. In urban areas, wireless communication infrastructures could be integrated into solar panels for architectural aesthetics or costs. Finally the integration of antennas and solar cell panels can be essential for the transportability of emergency autonomous communication stations.

This work focuses on the integration of SC into a particular type of antenna, namely the Reflectarray (RA) antenna. [9-10] Indeed, the flatness, low cost and high performances of RAs make them well suited for their integration into solar cell panels. Second, RA and SC are often found, or of interest, in the same platforms, such as satellites [7] or transportable autonomous communication systems [1, 11]. As illustrations, NASA has chosen a small spacecraft with RA and solar panel for a technology demonstration mission in 2014 (ISARA, Integrated Solar Array and Reflectarray Antenna). This satellite will deploy a panel covered on one side with SC and on other with a Ka-band RA. This setup spatially differentiates the two systems and therefore avoids detrimental interaction, such as the degradation of the solar cell efficiency. However, the direct integration of the RA on the solar cell could yield substantial additional flexibility (e.g. spinning satellites and deep space mission to other planets or L2 Lagrangien point of the earth-sun system). In terrestrial applications, the recent RESKUE program of the ESA [11] uses a deployable RA antenna for an autonomous and portable satellite communication station, which is another example where the integration of solar cell would be beneficial.

It is worth mentioning that the integration of RA and SC was previously considered in [12], where a full reflectarray of cross-dipoled element were etched onto a kapton substrate and laid on a traditional array of crystalline silicon solar cells. Even though the RA did form a coherent beam in the far field, the aperture efficiency of the antenna was only about 10%, far from the expected value of 40%. According to the authors, this was due to a lack of consideration and understanding of the RA substrate. Moreover, the kapton substrate reduced the solar cell efficiency by a very significant among, namely 40.5%. Therefore, this initial proof of concept came at the cost of unacceptable loss both at antenna and solar cell levels.

In this work, we show, thanks to careful characterization of the SC properties and detailed optimization of the RA element, that it is possible to achieve RA radiation efficiency while preserving good SC performances. The solar cells considered are (radiation hard) thin films of amorphous silicon while both metal and Transparent Conductive Oxides (TCOs) are investigated to implement the antenna conductors. Early developments related to this work were presented in the conference paper of [13].

This work was supported by the Swiss National Science Foundation (SNSF) under grant n°133583.

P. Dreyer and J. Perruisseau-Carrier are with the group for Adaptive MicroNano Wave Systems, LEMA/Nanolab, Ecole Polytechnique Fédérale de Lausanne (EPFL), CH-1015 Lausanne, Switzerland (e-mail: julien.perruisseau-carrier@epfl.ch).

M. Morales-Masis and C. Balif are with the Photovoltaics and thin-film electronics laboratory, Institute of Microengineering (IMT), Ecole Polytechnique Fédérale de Lausanne (EPFL), EPFL-STI-IMT-NE, Rue de la Maladière 71, CH-2000 Neuchatel, Switzerland.

S. Nicolay C. Balif are with the Centre Suisse d'Electronique et de Microtechnique (CSEM), CH-2002 Neuchâtel, Switzerland

## II. INTEGRATION CONCEPT

The principle of RA is well known [14-15]. In short, a RA is a reflector antenna in which the control of the phase of the reflected field is based on the control of the phase on the reflecting surface, while usual reflector as parabola uses the difference in the path length. In practice, the reflecting surface is achieved with a smart arrangement of elements with different phase shift.

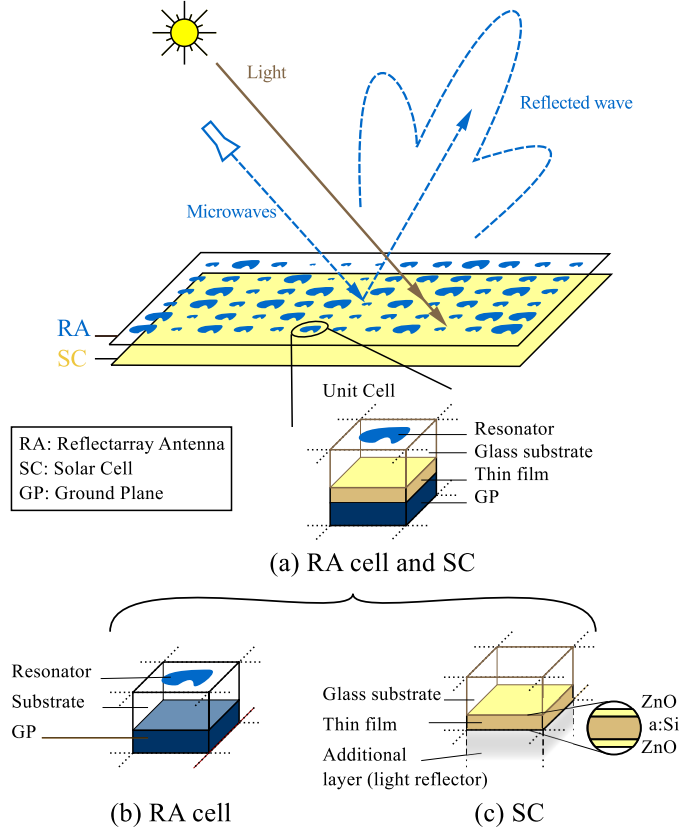

Fig. 1. General design of a RA integrated into solar cell.

Fig. 1 shows the basic principle of RA-SC integration proposed in this work, where an array of elements providing all phases in a 360° range are merged with a solar cell. Fig. 1(b) shows a typical RA cell made of a resonating patch element deposited on a substrate and terminated by a Ground Plane (GP). The geometry of the patch is used to control the phase of the cell. Concerning the SC, here we consider a single-junction thin film hydrogenated amorphous silicon (a-Si:H) solar cell, in the p-i-n configuration [16]. Though thin-film SC does not provide the highest efficiencies, they come with other benefits. In space for instance, first tests are putting in evidence their excellent radiation hardness, which is very profitable for long-term mission [17, 18]. Moreover, the efficiency is increased in space thanks to the richer blue content of the AM0 spectrum, and the high operating temperature allows a lower impact of the Staebler-wronski light induced degradation. Furthermore, they can easily be deposited on thin, lightweight and flexible substrates [16]. The flexibility of thin film solar cell simplifies the solar panel deployment and thus lightens the overall power system.

The thin film Si solar cell used in this work consists of a 2μm-thick ZnO:B front electrode deposited on a glass substrate, followed by a 250nm-thick hydrogenated amorphous silicon (a-Si:H) as the absorber layer, a 2.5 μm-thick ZnO:B as the back electrode and a metal back reflector. The sequence of the layers is shown in Fig. 1 (c). The ZnO:B front and back electrodes are deposited by low-pressure metalorganic chemical vapor deposition (LPCVD) and the a-Si:H layer by plasma enhanced-CVD (PECVD) [19].

The reflectarray integrated into a solar cell is a combination of the two previous structures as shown in Fig 1. (a) Thus, only a resonator and a ground plane need to be added to the standard SC structure to achieve the RA functionality. The concept is thus technologically and conceptually very simple. However, careful optimization must be done to achieve good RA performance while having the smallest impact of the efficiency of the solar cell, with numerous geometrical and material design variable involved

## III. SOLAR CELL CHARACTERIZATION AND SIMULATION

First the glass substrate and the solar cell described in section II were characterized. The glass was measured placing a sample into a rectangular waveguide at X-band (our frequency of interest) to extract permittivity and tangent loss from the measured scattering parameters, yielding $\varepsilon$ = 4.5 and $\tan\delta$ = 0.015.

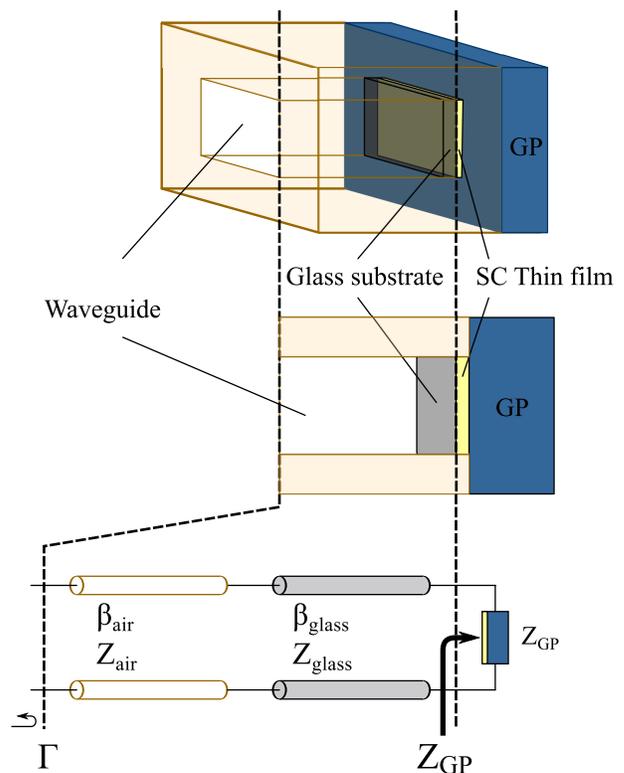

Fig. 2. Modelling of the cell in waveguide simulator and equivalent circuit model

The thin film SC selected in this work is only 4.75 μm thick, which is extremely thin regarding to the free-space wavelength at X-band frequency (thickness of the thin film is approximately $\lambda * 10^{-4}$). Since the thin film SC is pressed against a metallic plane Fig. 1(c), a pragmatic yet rigorous approach to our goal is to model the SC thin film over the GP as a single impedance $Z_{GP}$ 'looking' into the thin film, such as shown in Fig. 2. The value of the impedance was extracted from the reflected coefficient Γ measured at the extremity of a rectangular waveguide filled with the sample, i.e. a solar cell deposited on a glass substrate, and terminated with the GP. The extracted impedance $Z_{GP}$ is very close to that of an ideal ground plane, since at the design frequency 11.5 GHz $Z_{GP}$=0.5 + 1j Ω.

Both the glass and the effective ground plane are implemented in the finite element software Ansoft HFSS, to allow later optimization of the reflectarray cell. Accounting for the glass is trivial, while modelling the effective ground plane of impedance $Z_{GP}$ requires particular attention since such boundary condition in not readily available in the software. Nevertheless an exact representation can be achieved using the follow equivalence. Fig. 3 shows how a surface impedance Z followed by a virtual waveguide that is terminated with a short cut can simulate $Z_{GP}$ in HFSS.

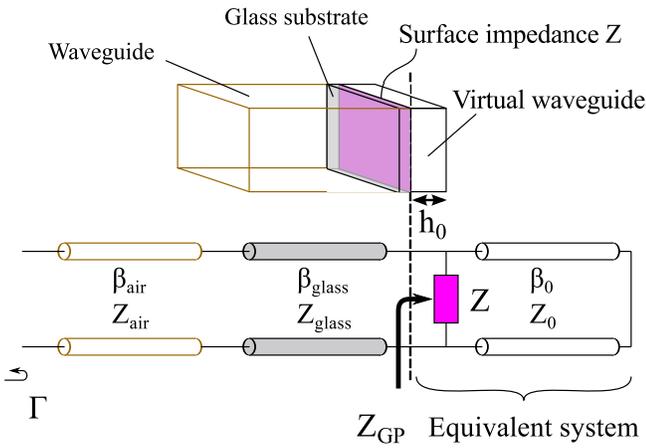

Fig. 3. Numerical design of the solar cell over the ground plane in a rectangular X-Band waveguide and its equivalent transmission line schema.

Indeed, the input impedance $Z_{GP}$ is equivalent to the impedance Z connected in parallel with a shorted transmission line:

$$\frac{1}{Z_{GP}} = \frac{1}{Z} + \frac{1}{jZ_0 \tan(\beta_0 h_0)} \quad (1)$$

The impedance Z of given transmission line ($Z_0$, $\beta_0$, $h_0$) in the equivalent system of $Z_{GP}$ is then simply given by

$$Z = \frac{1}{\frac{1}{Z_{GP}} - \frac{1}{jZ_0 \tan(\beta_0 h_0)}} \quad (2)$$

The impedance Z is frequency dependent because it depends on the propagation constant of the virtual transmission line $\beta_0$. The validity of the approach can easily be verified by comparing results obtained by simulation of the equivalent system and the desired corresponding analytical solution. A perfect agreement was obtained here, and thus the HFSS model can now be used for the optimization of the reflectarray cell itself.

## IV. RA Cell Optimization

The RA Cell developed in this work is presented in Fig. 4 and is composed of four primary layers.

- **Resonator:** On the top, a square-shaped ring patch resonator of dimensions $w_p$ and $a_p$ is deposited on a glass substrate. It is a well-known performing topology for RA cells, which is also favourable for minimal optical blockage. The optical transparency and microwave sheet resistance of the conductive top layer are fundamental since they impact on the RA radiation efficiency and SC efficiency. However transparency and conductivity of conductors are antinomic and it cannot be guessed a-priori if a large transparent ring patch (but with poor conductivity) is better than a thin ring patch of a good conductor (but not transparent), since both thickness and surface of the used conductive material are variables in the design. Therefore both standard copper and Transparent Conducting Oxides (TCOs) are considered next.
- **Glass substrate:** The glass substrate is a commercial product used for solar cell deposition and is provided by Schott AG (AF32). Its electrical properties were measured earlier (ε=4.5 and tδ=0.015).
- **Solar cell:** The thin film a-Si:H SC is deposited on the other side of the glass substrate.
- **Ground plane:** The cell is terminated with a metallic reflector.

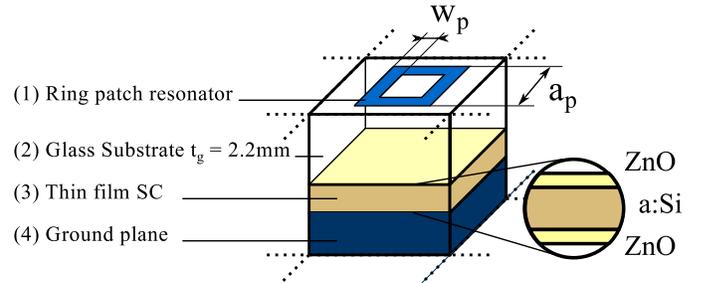

Fig. 4. Complete structure of the RA cell integrated into a thin SC of amorphous silicon.

It is important to note that a *single 'design solution'* in our optimization problem consists of a *set* of reflectarray cells. This is simply because a full reflectarray is built from a family of cells providing all required phases, ideally 360°. In this work, the phase control within a given solution is achieved by varying the size of the ring ($a_p$). Fig. 5(a) schematically shows how the size of the patch ($a_p$) influences the phase response of elements for a design solution.

We wish to design a solution, i.e. a set of cells, having both good microwave and solar cell performances. To this purpose the following 'metrics' are considered:
- Phase range
- Average magnitude of the reflected wave $\tilde{\rho}$
- Average transparency $\tilde{T}$

The quantities $\tilde{\rho}$ and $\tilde{T}$ are averaged over the family of cells providing all phases. However to provide a fair representation of the eventual array performance in terms of microwave loss and optical transparency, this averaging must be done as follows. First, and targeting generality, we assume that the complete RA antenna system is composed of a set of cells providing all phases *with equal probability* (if the actual probability distribution of phases on the array is known it can also very easily be taken into account). Therefore the way to compute $\tilde{\rho}$ and $\tilde{T}$ for a solution at a given frequency is to average them over a finite set of cells corresponding to a uniform phase distribution within the available range. This is better understood by a simple observation of Fig. 5.

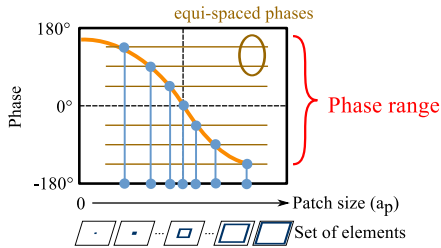

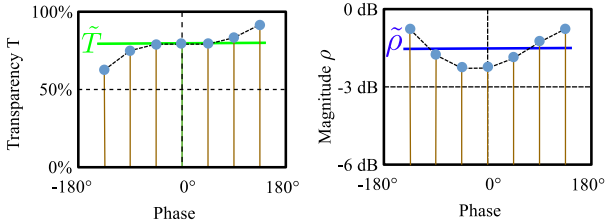

Fig. 5. Phase response (a), transparency (b) and magnitude of reflection (c) of a design solution for a set of elements providing phases with uniform distribution.

Next, some technological parameters are fixed. First, for the reason explained above, both copper and TCO are considered for the reflectarray ring resonator. Since the sensitivity of the solar cell is limited to the visible spectrum (i.e. form 400nm to 700nm), we have measured transparency of both glass and TCO layer on glass at this frequency range as shown in Fig. 6. Transparency of 92% for the glass and 78% for the TCO on glass were found in the visible spectrum (400nm-700nm) whereas transparency of the copper is logically of 0%. Thus, the overall optical transparency (in the visible spectrum) of a RA cell depends on the filling factor of the resonator and the transparencies measured above. Previous materials affect also RA performances since the RA functionality depends on sheet resistance of the rings, which is fixed here to Rs = 0.027 Ω/sq for copper and Rs = 2 Ω/sq for TCO.

Second, the substrate thickness was fixed to $tg$ = 2.2 mm so as to comply with existing solar cell glass thicknesses readily available for this work.

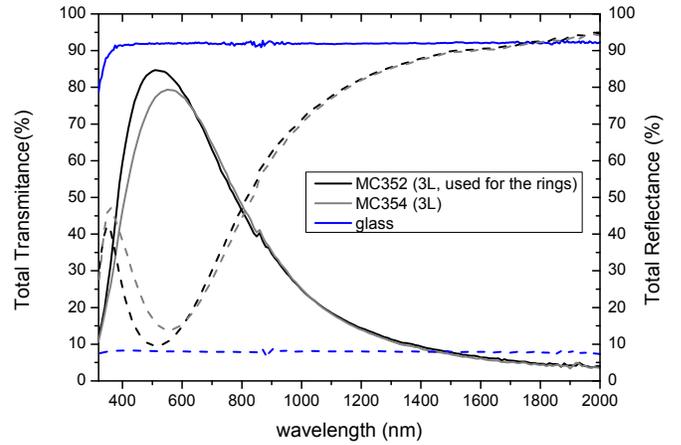

Fig. 6. Total transmittance (transparency) of the glass substrate and the TCO layer deposited on the glass substrate (MC352).

Fig. 7-8 show the results obtained for solutions (set of elements where the ring size varies from $a_p$ = 0mm to $a_p$ = 10mm) corresponding to different ring widths $w_p$, and discarding the ones exhibiting a phase range smaller than 270°. Indeed, smaller ranges are of moderate interest for real reflectarray implementations, while such a value already allows close to negligible degradation on the eventual array patterns [20].

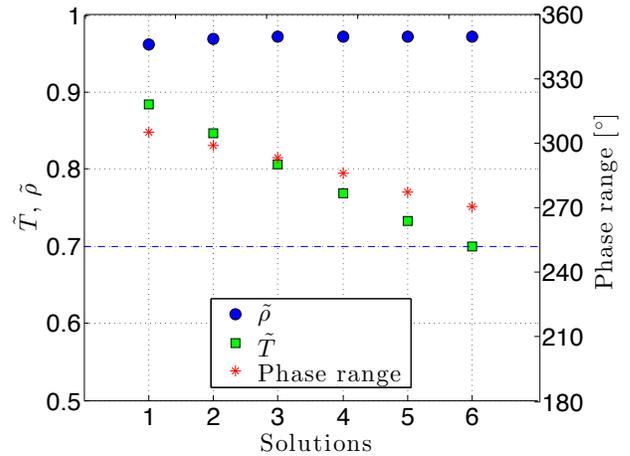

Fig. 7. **Copper**: Results for cells with glass thickness of 2.2mm and for patches where $w_p$ increases from 0.5mm (Solution 1) to 3mm (Solution 6).

As expected, there is not a unique optimal solution since some solutions present better microwave loss than other while having poorer optical transparency, and vice-versa. Therefore an optimal solution would depend on the application requirement. For instance, for the best microwave antenna performance one should select Solution 6 in Fig 7, where the average reflection magnitude $\tilde{\rho}$ reaches 0.97 (-0.25 dB). However, the SC will not work at its maximum efficiency because the average transparency $\tilde{T}$ is only of 70%. Therefore, this solution is interesting for high efficiency antennas where the SC efficiency is of less importance.

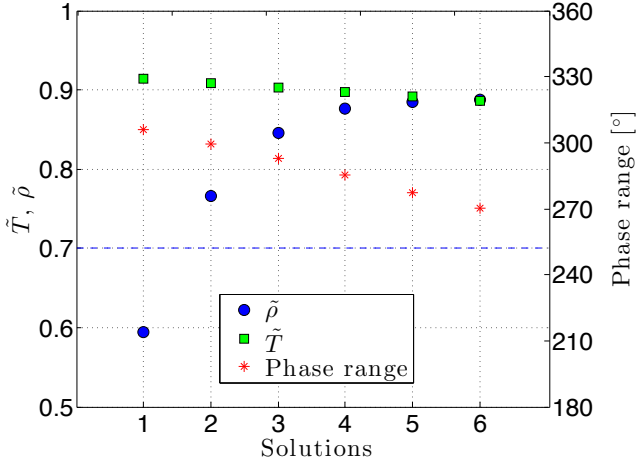

Fig. 8. **TCO:** Results for cells with glass thickness of 2.2mm and for patches where $w_p$ increases from 0.5mm (Solution 1) to 3mm (Solution 6).

If one rather needs a design that maximizes the solar cell efficiency, Solution 1 in Fig. 8 should be selected, where the average optical transparency reaches 91%. In this case the antenna efficiency will be poorer since it reaches ($\tilde{\rho} = 0.6 = $ -4.44dB). In practice, intermediate solutions might obviously be selected.

## V. EXPERIMENTAL RESULTS

Solution 2 in Fig. 7 and solution 4 in Fig. 8 were selected for experimental validation, and will referred hereafter as Solution A (Copper) and Solution B (TCO). The ring width $w_p$ (not to be confused with the ring size, see Fig. 4) is 1 mm for Solution A and 2mm for Solution B. Four patch sizes ($a_p$) were selected to be representative of the different phases and losses within each solution family. More precisely, selected patches widths are $a_p$=2mm, 4mm, 5mm and 8mm for TCO and $a_p$=2mm, 4.25mm, 5mm for cooper.

The detailed simulated performances of the selected solutions are shown in Fig. 9-12 for normal incidence (PEC-PMC waveguide). Fig. 9-10 show the magnitude and the phase of the reflected wave for both solutions.

Fig. 11-12 show transparency and magnitude of the reflected wave of solutions A and B over the phase range at a given frequency. As expected from Fig. 7-8, Solution B (TCO) has good optical performances and reach 90% of average transparency and -1.15dB of average magnitude of the reflected wave. Solution A is more suitable for an antenna with good performance since it reach -0.25dB for $\tilde{\rho}$, while the average transparency reduces to 85%.

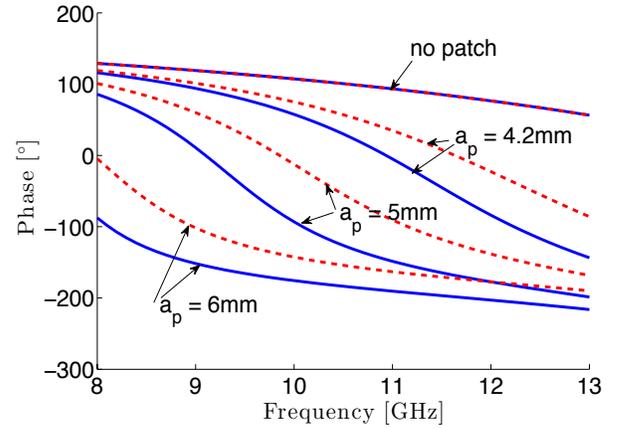

Fig. 9. Phase reflection of Solution A (Blue) and Solution B (Red) for different ring patches ($a_p$).

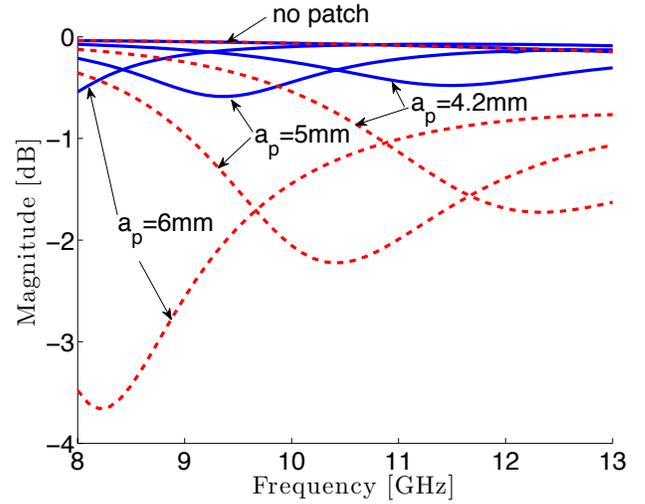

Fig. 10. Magnitude reflection of Solution A (Blue) and Solution B (Red) for different ring patches ($a_p$).

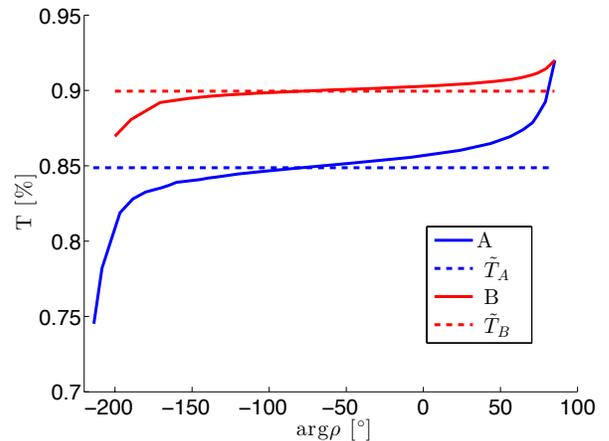

Fig. 11. Optical transparency (in the visible spectrum) for the design solutions A and B calculated for a cell working at 11.5 GHz. The graph shows the quantities of interest as a function of the reflection phase for solution A, and B.

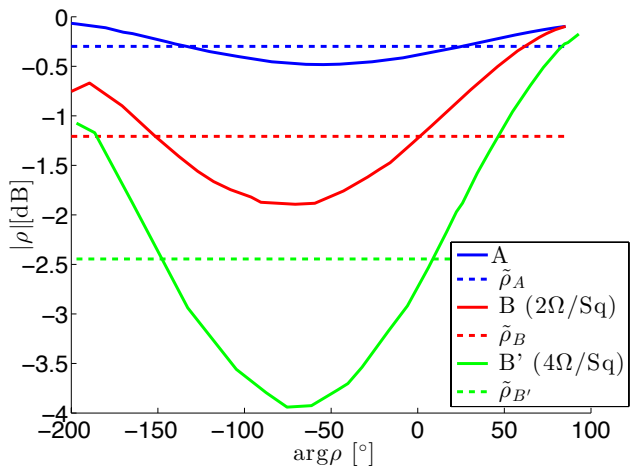

Fig. 12. Microwave loss for the design solutions A and B calculated for a cell working at 11.5 GHz. The graph shows the quantities of interest as a function of the reflection phase for solution A, and B.

The selected representative cells were then fabricated and measured. For convenience, patches were deposited on a 1.1mm glass substrate while the solar cell was deposited on another 1.1mm glass substrate. The transparent conductive rings were sputtered onto the glass substrates with the use of a shadow mask for patterning. These rings consist of a multilayer stack of metal oxide (MO) / metal / metal oxide (MO). A very thin metal layer is introduced between the two MO layers to reduce the resistance to the required value of 2 ohm/sq, while the MO layers reduce the reflectance of the thin layer and improve the overall transmittance of the stack [21-22].

The RA cell structure is obtained by superposing the two pieces of glass. A picture of the different building blocks is shown in Fig. 13.

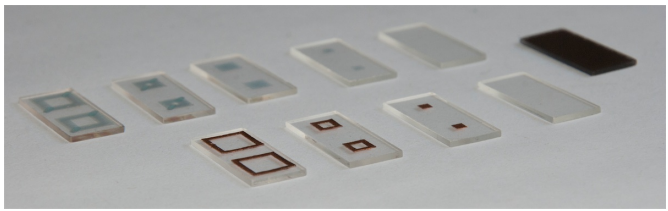

Fig. 13. Sample of RA cell. Top: TCO. Bottom: Copper. Top right: Solar Cell.

The cells are measured in a RWG simulator and compared to simulations in the same setup. Corresponding results are shown in Fig. 14-15 for the copper and Fig. 16-17 for the TCO, respectively. A good agreement is observed between simulation and measurement. However it was found by fitting that the actual microwave surface resistance of the implemented TCO was about 4 Ω/sq, whereas 2 Ω/sq (DC-value) had been considered for optimization. Therefore, to assess the related performance degradation we recomputed the expected solution family performance in the case of a sheet resistance of 4 Ω/sq, and corresponding results are reported in Fig. 12 in comparison with the 2 Ω/sq case. Though obviously the average microwave loss is higher than initially expected, reaching now $\tilde{\rho} = -2.45$ dB, this value is still the best demonstrated so far and very encouraging for real application, particularly since it comes with very good optical average transparency ($\tilde{T} = 90\%$ in the visible range of the spectrum, i.e. from 400nm to 700nm). If such microwave loss is not acceptable, copper should be selected to achieve very low loss, at the cost of some reduction in optical transparency. Obviously, any technological improvement in the TCO surface resistance at a given optical transparency will allow improved performance.

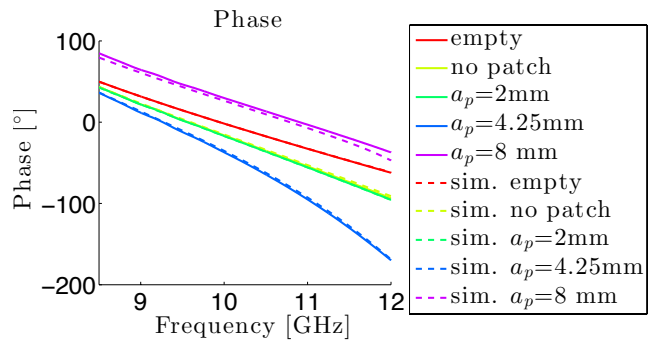

Fig. 14. Solution A: **Copper**. Phase response of solution A for different patch sizes.

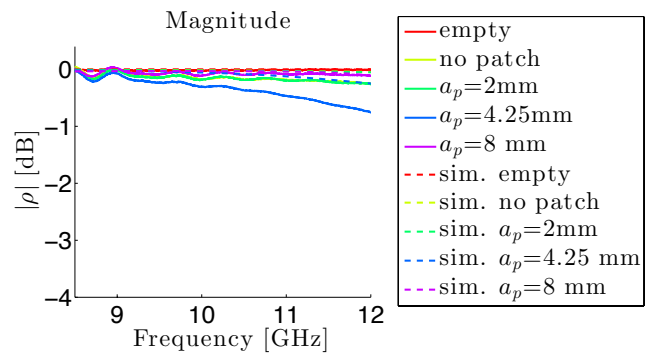

Fig. 15. Solution A: **Copper**. Magnitude response of solution A for different patch sizes.

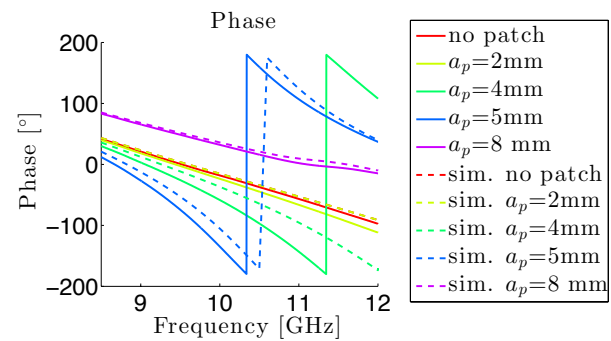

Fig. 16. Solution B' (4 Ω/sq): **TCO**. Phase response of solution B' for different patch sizes.

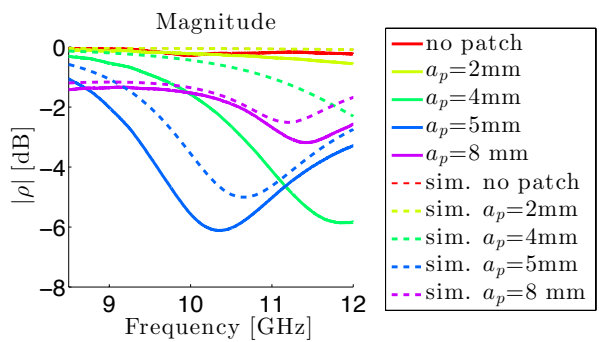

Fig. 17 Solution B' (4 Ω/sq): **TCO**. Magnitude response of solution B' for different patch sizes.

## VI. Conclusion

The viability of using solar cell panels as reflectors in reflectarray antenna configurations has been demonstrated. A detailed procedure for optimizing reflectarray cells family has been developed and successfully applied. The results are validated by measurement of fabricated cells using both standard copper and transparent conductors. Interestingly, transparent conductors are only useful when solar cell efficiency is the main requirement. For very good microwave performance at slightly degraded optical transparency, it is preferable to use very thin patterns of opaque standard conductor. These results open the path to further advances in the integration of reflectarrays on solar cell panels.

### Acknowledgement

P. Dreyer thanks Dr. Juan Sebastián Gómez Díaz and Dr. Tomislav Debogović for their help and support during this work.

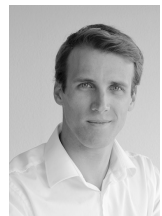

**Philippe Dreyer** was born in Neuchatel, Switzerland in 1986. He received the B.S. and M.S degrees in microengineering from the Ecole Polytechnique Fédérale de Lausanne, Lausanne, Switzerland, in 2012.

In 2013, he did an internship as research assistant with the group for Adaptative Micronano Wave Systems, Lausanne Switzerland. His research interest includes the development of reflectarray antennas integrated in solar cell using metallic and transparent conductors.

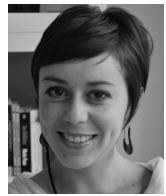

**Monica Morales-Masis** was born in Costa Rica in 1982. She received her Ph.D. degree in Physics from Leiden University, The Netherlands in January 2012. The same year she joined the Photovoltaics and Thin-Film Electronics Laboratory, Ecole Polytechnique Fédérale de Lausanne, Neuchatel, Switzerland. Nowadays she is Team Leader of the Transparent Conductive Oxides (TCO) group. Her research interests include the fundamental investigation, fabrication and application of transparent conductive oxides in optoelectronic devices, flexible and transparent electronics.

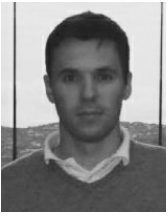

**Sylvain Nicolay** obtained his thesis in 2008 on MOCVD growth of nitride based heterostructures and worked in the PVLab on the modelling/in-depth understanding of TCO growth and transport mechanisms. Dr. Nicolay is now in charge of coatings for the PV center at CSEM.

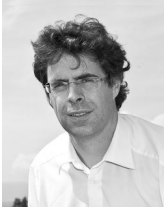

**Christophe Ballif** received the M.Sc. and Ph.D. degrees in physics from the Ecole Polytechnique Fédérale de Lausanne (EPFL), Switzerland, in 1994 and 1998, respectively, focusing on novel photovoltaic materials.

In 2004, he became a Full Professor and Chair with the Institute of microengineering, University of Neuchâtel, Switzerland. In 2009, the Institute of Microengineering was transferred to EPFL. He is the Director of the Photovoltaics and Thin-Film Electronics Laboratory within the Institute, and since 2013, he has also been the Director of the PV-Center within the Swiss Center for Electronics and Microtechnology (CSEM), Neuchâtel. His research interests include thin-film silicon, high-efficiency heterojunction crystalline cells, module technology, thin film for electronic applications, contributing to technology transfer, and industrialization of novel devices.

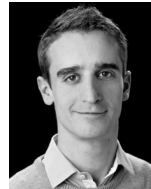

**Julien Perruisseau-Carrier** (S'07-M'09-SM'13) received the M.Sc. and Ph.D. degrees from the Ecole Polytechnique Fédérale de Lausanne (EPFL), Lausanne, Switzerland, in 2003 and 2007, respectively.

In 2003, he was with the University of Birmingham, UK. From 2004 to 2007, he was with the Laboratory of Electromagnetics and Acoustics (LEMA), EPFL, where he completed his PhD while working on various EU funded projects. From 2007 to 2011 he was with the Centre Tecnològic de Telecomunicacions de Catalunya (CTTC), Barcelona, as an associate researcher. Since June 2011 he is a Professor at EPFL, where he leads various projects and work packages at the National, European Space Agency, European Union, and industrial levels. His main research interest concerns interdisciplinary topics related to electromagnetic waves from microwave to terahertz: dynamic reconfiguration, application of micro/nanotechnology, joint antenna-coding techniques, and metamaterials. He has authored +80 and +50 conference and journal papers in these fields, respectively.

Julien Perruisseau-Carrier was the recipient of the Raj Mittra Travel Grant 2010 presented by the IEEE Antennas and Propagation Society, and of the Young Scientist Award of the URSI Intern. Symp. on Electromagnetic Theory, both in 2007 and in 2013. He currently serves as an Associate Editor of the IEEE TRANSACTIONS ON ANTENNAS AND PROPAGATION..